\documentclass{article}%
\usepackage[colorlinks,linkcolor=black,citecolor=black,urlcolor=black]%
{hyperref}
\usepackage{amsmath}
\usepackage{amssymb}
\usepackage[conf]{ametsoc}
\usepackage{graphicx}
\usepackage{amsfonts}%
\providecommand{\U}[1]{\protect\rule{.1in}{.1in}}

\renewcommand{\cite}[2][]{\citep[#1]{#2}}
\begin{document}
\title{A wildland fire modeling and visualization environment\footnotemark[1]}
\author{Jan Mandel\footnotemark[2], Jonathan D. Beezley\footnotemark[2],
Adam K. Kochanski\footnotemark[3], \\
Volodymyr Y. Kondratenko\footnotemark[2], Lin Zhang\footnotemark
[4], Erik Anderson\footnotemark[4], \\
Joel Daniels II\footnotemark[5], Cl{\'a}udio T. Silva\footnotemark
[4], and Christopher R. Johnson\footnotemark[4]\\
\ \\
\footnotemark
[2] \ Department of Mathematical and Statistical Sciences, University of Colorado Denver \\
\footnotemark[3] \ Department of Atmospheric Sciences, University of Utah \\
\footnotemark
[4] \ Scientific Computing and Imaging Institute, University of Utah\\
\footnotemark[5] \ Polytechnic Institute, New York University}
\maketitle
\begin{abstract}
We present an overview of a modeling environment, consisting of a coupled atmosphere-wildfire model,
utilities for visualization, data processing, and diagnostics, open source software repositories, and a community
wiki. The fire model, called SFIRE, is based on a fire-spread model, implemented by the level-set method,
and it is coupled with the Weather Research Forecasting (WRF) model.
A version with a subset of the features is distributed with WRF 3.3 as WRF-Fire. In each time step, the
fire module takes the wind as input and returns the latent and sensible heat fluxes. The software architecture
uses WRF parallel infrastructure for massively parallel computing. Recent features of the code include
interpolation from an ideal
logarithmic wind profile for nonhomogeneous fuels and ignition from a fire perimeter with an atmosphere and
fire spin-up. Real runs use online sources for fuel maps, fine-scale topography, and meteorological data, and can
run faster than real time.
Visualization pathways allow generating images and animations in many packages, including VisTrails, VAPOR,
MayaVi, and Paraview, as well as output to Google Earth. The environment is available from \url
{openwfm.org}.
New diagnostic variables were added to the code recently, including a new kind of fireline intensity,
which takes into account also the speed of burning, unlike Byram's fireline intensity.
\end{abstract}
\footnotetext[1]{Paper 6.4, Ninth Symposium on  Fire and Forest Meteorology,
Palm Springs, CA, American Meteorological Society, October 2011.
Available at \mbox{\protect\url
{http://ams.confex.com/ams/9FIRE/webprogram/Paper192277.html}}.
This research was supported by NSF grant AGS-0835579 and by NIST Fire Research Grants Program grant 60NANB7D6144.
}

\section{Introduction}

A model that can be used by others than the authors requires an extensive
infrastructure far beyond the simulation code itself. Social aspects are just
as important as technical ones, if the model is to be used by humans. For this
reason, the OpenWFM.org environment consists of multiple components, many
under continued development:

\begin{itemize}
\item The wildland fire simulation code SFIRE\ coupled with a mesoscale
atmospheric simulation code, the Weather Research and Forecasting (WRF) model
(Sec.~\ref{sec:model}).

\item Facilities for acquisition of data from online databases, and
preprocessing of fine-resolution data (described separately in
\citet{Beezley-2011-IHS}; see also \citet{Beezley-2011-IHD}).

\item Documentation and diagnostic utilities (Sec.~\ref{sec:documentation})

\item Visualization, clients, web portals, and social networks
(Sec.~\ref{sec:visualization})
\end{itemize}

\section{Coupled fire-atmosphere model}

\label{sec:model}

\subsection{WRF coupled with fire spread model}

Wildland fire is a complicated multiscale process. Fortunately, a practically
important range of wildland fire behavior can be captured by the coupling of a
mesoscale weather model with a simple 2D fire spread model
\cite{Clark-1996-CAF-x,Clark-1996-CAM,Clark-2004-DCA,Coen-2005-SBE}. Weather
has a major influence on wildfire behavior; in particular, wind plays a
dominant role in the fire spread. Conversely, the fire influences the weather
through the heat and vapor fluxes from burning hydrocarbons and evaporation.
The buoyancy created by the heat from the fire can cause tornadic strength
winds, and the wind and the moisture from the fire affect the atmosphere also
away from the fire. It is well known that a large fire \textquotedblleft
creates its own weather.\textquotedblright

\begin{figure}[t]
\begin{center}
\includegraphics[width=7in]{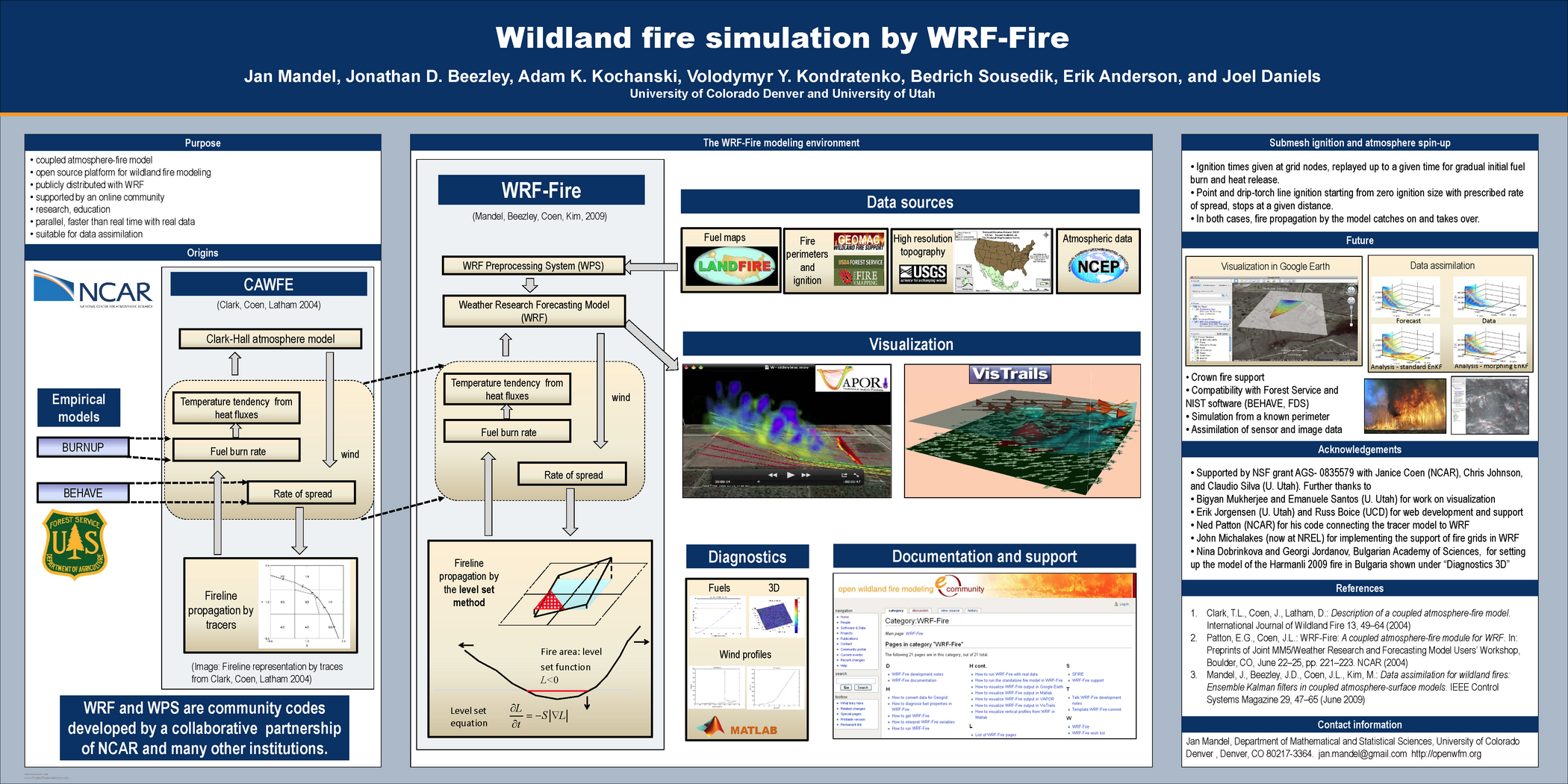}
\end{center}
\caption{An overview of the wildfire simulation environment
\cite{Mandel-2010-WFS}.}%
\label{fig:poster}%
\end{figure}

The code combines the Weather Research and Forecasting Model (WRF) with the
semi-empirical fire spread model \cite{Rothermel-1972-MMP}, with the fire
spread implemented by the level-set method. Alternatives to the Rothermel's
model are in progress. The model is able to run faster than real time on
several hundred cores, with the fire model resolution of few meters and
horizontal atmospheric resolution on the order of 100m, for a large real fire
\cite{Jordanov-2011-SHF}.

\citet{Kochanski-2010-EFP} compared simulation results with measurements on
the FireFlux grass fire experiment \cite{Clements-2007-ODW}.
\citet{Dobrinkova-2011-WAB} simulated a fire in Bulgarian mountains using real
meteorological and geographical data, and ideal fuel data.
\citet{Beezley-2010-SMC} simulated a fire in Colorado mountains using real
data from online sources. \citet{Jordanov-2011-SHF} simulated a large fire in
Bulgaria with real data, including a fuel map derived from satellite
mesurements, and compared the results with the actual fire area.

Further details, references, and acknowledgements can be found in
\citet{Mandel-2009-DAW,Mandel-2011-OCA,Mandel-2011-CAF}. See also
Fig.~\ref{fig:poster} for an overview poster.

\subsection{Related work}

Wildland fire models range from tools based based on fire spread rate formulas
\cite{Rothermel-1972-MMP,Rothermel-1983-HTP}, such as BehavePlus
\cite{Andrews-2007-BFM} and FARSITE \cite{Finney-1998-FFA}, suitable for
operational forecasting, to sophisticated 3D computational fluid dynamics and
combustion simulations suitable for research and reanalysis, such as FIRETEC
\cite{Linn-2002-SWB} and WFDS \cite{Mell-2007-PAM}. BehavePlus, the PC-based
successor of the calculator-based BEHAVE, determines the fire spread rate at a
single point from fuel and environmental data; FARSITE uses the fire spread
rate to provide a 2D simulation on a PC; while FIRETEC and WFDS are physically
more accurate and run much slower than real time.

The level set-method was used for a surface fire spread model in
\citet{Mallet-2009-MWF}. \citet{Filippi-2009-CAF} coupled the atmospheric
model, Meso-nh, with fire propagation by tracers. Tiger \cite{Mazzoleni-T2F}
uses a 2D combustion model based on reaction-convection-diffusion equations
and a convection model to emulate the effect of the fire on the wind. FIRESTAR
\cite{Morvan-2004-MPW} is a physically accurate wildland fire model in two
dimensions, one horizontal and one vertical. UU LES-Fire \cite{Sun-2009-IFC}
couples the University of Utah's Large Eddy Simulator with the tracer-based
code from CAWFE. See the survey by \citet{Sullivan-2009-RWF} for a number of
other models.

\subsection{New features}

Recent features of the code include interpolation from an ideal logarithmic
wind profile for nonhomogeneous fuels \cite{Mandel-2011-CAF}, important for
simulation at a variety of scales, and ignition from a fire perimeter with an
atmosphere and fire spin-up \cite{Kondratenko-2011-IFP}. The perimeter
ignition is important for practical application and the technique will be
adapted also for data asimilation, where the atmosphere state needs to be
adjusted when the fire state changes due to new data.

New features which were added to the code since the paper
\citet{Mandel-2011-CAF} include computation of fireline intensities. Byram's
fireline intensity \cite{Byram-1959-CFF} is the heat produced per unit length
of the fireline in unit time (J/m/s) in the so-called flaming zone behind the
fireline. Hence, it is given by%
\begin{equation}
I=HRw \label{eq:byram-intensity}%
\end{equation}
where $H$ (J/kg) is the heat contents of the fuel, $R$ (m/s) is the spread
rate, and $w$ (kg/m$^{2}$) is the fuel amount that burns in the flaming zone.
In practice, the fuel amount burned $w$ is estimated as a fixed fraction of
the fuel load $w_{0}$ (kg/m$^{2}$).

Though Byram's fireline intensity is routinely used for practical guidance,
\emph{it does not depend on the speed of burning}. However, if the fuel burns
slowly, much of the burning takes place at a distance from the fireline and
thus may not contribute much to the severity of the fire, while a fast burning
fuel will release its heat close to the fireline.

For this reason, we introduce \emph{a new concept of fireline intensity as the
amount of heat generated by the advancing fireline from the newly burning fuel
only, in a small unit of time}. Assume that the fuel fraction after ignition
decreases exponentially with the time $t$ from ignition, as
\[
e^{-\tau/T_{\mathrm{f}}},
\]
where $T_{\mathrm{f}}$ is the fuel burn time, i.e., the time when
$1-e^{-1}\approx64\%$ of fuel has burned. Straightforward application of
calculus then shows that the new fireline intensity is
\begin{equation}
J=\frac{HRw_{0}}{2T_{\mathrm{f}}}\text{ (J/m/s}^{2}\text{).}
\label{eq:new-intensity}%
\end{equation}

Unlike Byram's fireline intensity (\ref{eq:byram-intensity}), the new fireline
intensity, given by (\ref{eq:new-intensity}), takes into account the effect
that a faster burning fuel will create a more intense heat concentrated at the
fireline. The reason why the time unit is squared is that over a longer time,
the fireline advances over a longer distance, and the fuel it advances over
also has longer to burn.

The code computes the fireline intensities and the reaction intensity (which
is the same as the released heat flux intensity, J/s/m$^{2}$) for the
simulated fire to estimate its severity. The fireline intensities are computed
from the fire rate of spread $R$. Since $R$ is well-defined on the fireline
only, the fireline intensities are defined on mesh nodes next to the fireline
only as well.

Separate computations are made as a component of a fire danger rating to
estimate the severity of a potential fire, that is, to answer the question: if
a fire breaks out, how bad would it be? These quantities are computed from the
maximal rate of spread at any location for the wind speed and the slope at
that location, and they can be used to plot potential fire severity maps.

\section{OpenWFM wiki}

\label{sec:documentation}

\begin{figure}[t]
\par
\begin{center}
\includegraphics[width=7in]{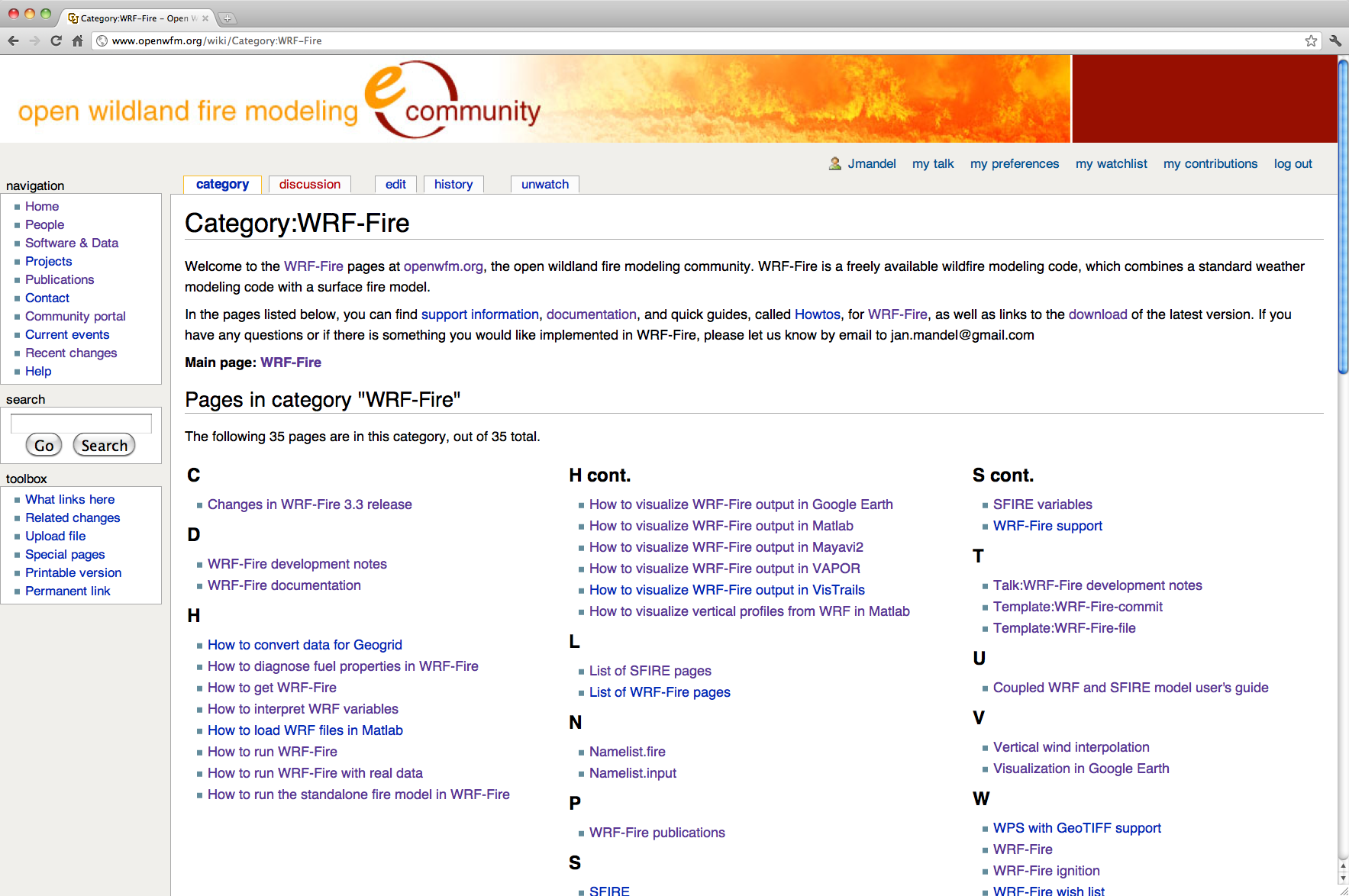}
\end{center}
\caption{Support and documentation pages on the OpenWFM wiki at
{\url{http://www.openwfm.org/wiki/List_of_SFIRE_pages}}}%
\label{fig:openwfm}%
\end{figure}

\begin{figure}[pt]
\par
\begin{center}
\includegraphics[width=7in]{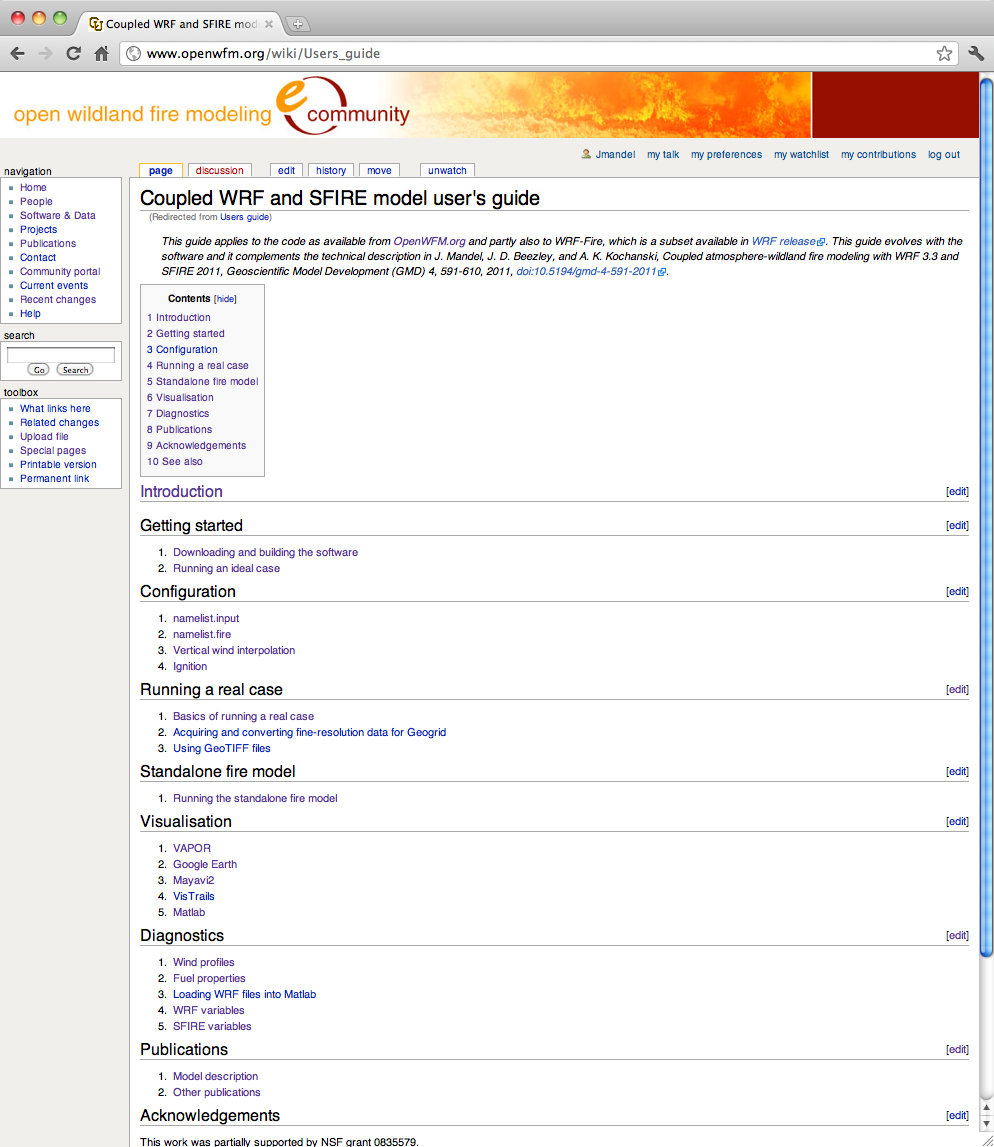}
\end{center}
\caption{User's Guide front page at
{\url{http://www.openwfm.org/wiki/Users_guide}}}%
\label{fig:users_guide}%
\end{figure}

Fig.~\ref{fig:openwfm} shows the list of all wiki pages for the support of the
simulation code. The pages starting with \textquotedblleft How
to\textquotedblright\ are short guides that allow the user do basic things
simply. Users can then investigate variations at their leisure.

Diagnostic utilities, described in the wiki and available from source code
repositories, allow the user, e.g., generate graphs of fire spread rate as a
function of wind speed to aid in debugging fire properties, or draw wind
profiles at a specified point (the wind speed as a function of height).

Since most the user's guide (appendix in \citet{Wang-2010-AUG}) was extracted
from the wiki articles anyway, and the wiki has much more information, such as
the diagnostic utilities, the existing wiki articles were now organized into a
new, continously updated User's Guide at
\url{http://www.openwfm.org/wiki/Users_guide} (Fig.~\ref{fig:users_guide}).
This required adding only a minimal amount of material.

The OpenWFM\ wiki provides a social environment, where users can edit and
comment on each other's work. It is using the WikiMedia software.

\section{Visualization, clients, web portals, and social networks}

\label{sec:visualization}

\subsection{Google Maps and Google Earth}

\begin{figure}[t]
\begin{center}
\includegraphics[width=7in]{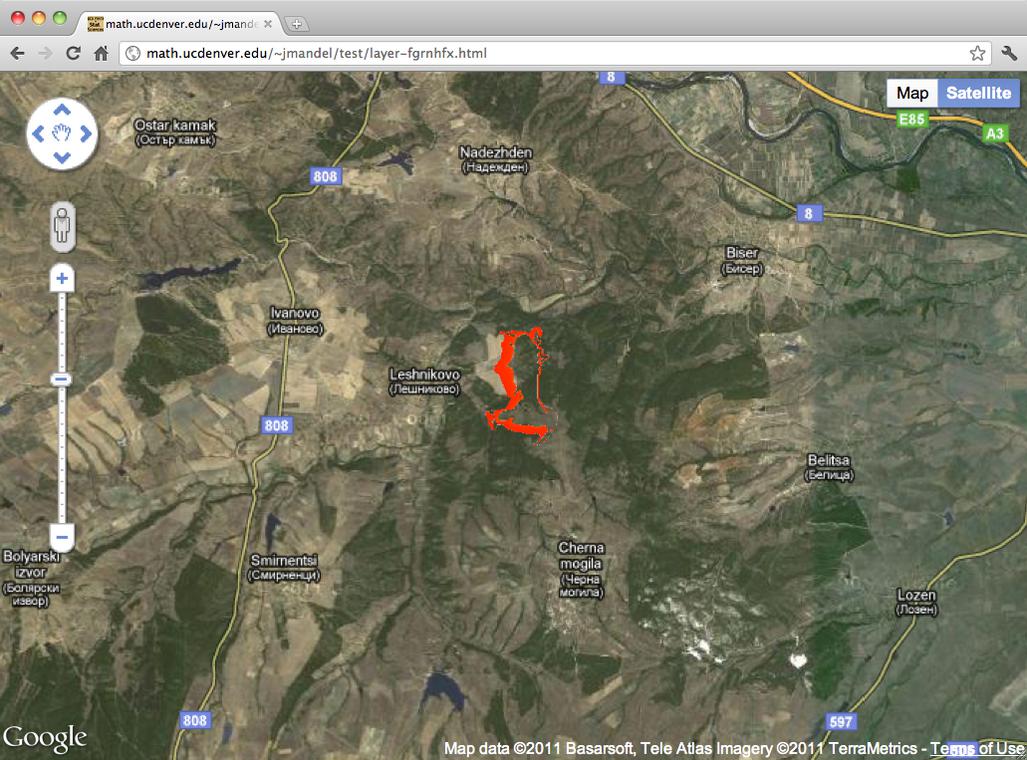}
\end{center}
\caption{Visualization in Google Maps web portal, 2009 Harmanli fire,
Bulgaria. Simulation data from \citet{Jordanov-2011-SHF}.}%
\label{fig:google_maps}%
\end{figure}

\begin{figure}[t]
\begin{center}
\includegraphics[width=7in]{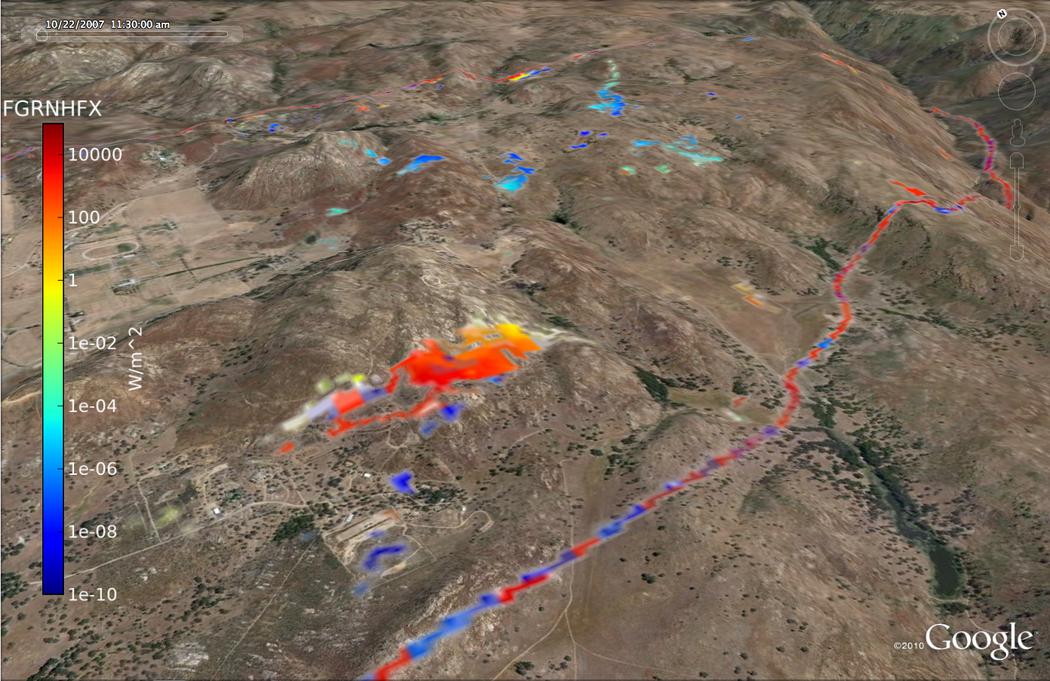}
\end{center}
\caption{Visualization in Google Earth client, 2010 Mt Carmel Fire, Israel.}%
\label{fig:google_earth}%
\end{figure}

\begin{figure}[t]
\begin{center}
\includegraphics[width=7in]{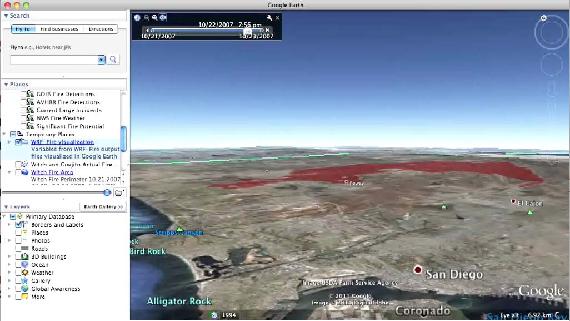}
\end{center}
\caption{Visualization in Google Earth client, 2007 Witch fire, CA.}%
\label{fig:google_earth_movie}%
\end{figure}

We have proposed Google Earth visualization already in
\citet{Douglas-2006-DVW}. Since then, Google API\ has become a de-facto
standard in wildland fire visualization. Both Google Earth and Google Maps use
the same file format, KML (Keyhole Markup Language). We use the compressed
format, KMZ. A KMZ file is a zipped directory with XML code containing drawing
instructions and metadata, such as geolocation, and (optionally) image files
to be displayed on the Earth surface. We use the image files to display the
heat flux (Figs.~\ref{fig:google_maps} and \ref{fig:google_earth})\ or fire
area (Fig.~\ref{fig:google_earth_movie}). We render the file in Google Maps by
posting it on the web and providing its URL\ to the Google Maps Javascript
API, in a web page (Fig.~ \ref{fig:google_maps}), or loading the file into the
Google Earth application. Google Earth can animate a KML\ file with multiple
images as frames. Animation in Google Maps requires custom Javascript coding
and is under development. Also, surface images to be visualized in Google
Earth and Maps API\ use a different map projection than WRF, so the mesh nodes
in WRF\ and the pixels in the image do not line up, and the simulation values
need to be interpolated.

Future extensions include a web portal or a client application based on Google
Maps, which can initiate a fire simulation on a cluster.

\subsection{VisTrails and CrowdLabs}

\begin{figure}[t]
\begin{center}
\includegraphics[width=7in]{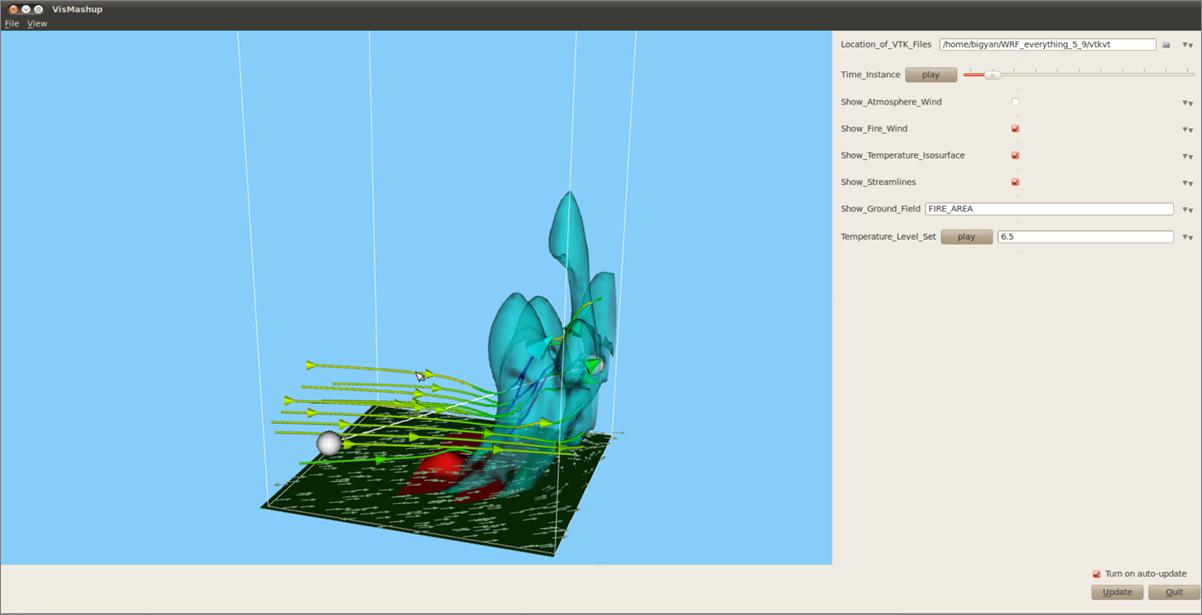}
\end{center}
\caption{Visualization in VisTrails client.}%
\label{fig:vistrails}%
\end{figure}

VisTrails (\citet{Freie-2011-V}, \url{http://www.vistrails.org}) is a
scientific development environment for workflow and provenance management. It
transparently stores user interactions within the system to ensure
reproducibility of the produced data sets and visualizations. VisTrails
provides a graphical programming interface that was specifically tailored for
the generation of custom scientific visualizations. Through a drag-and-drop
interface, users may interconnect modules that are responsible for reading a
file, parsing the data, and performing different visualization tasks.
VisMashups \cite{Santos-2009-VSC} is a framework for the automatic development
of simplified user interfaces from annotated workflows developed within the
VisTrails environment.

Together, VisTrails and VisMashups, provide the tools to generate customizable
visualizations for coupled WRF and fire simulations. Demonstrated in
Fig.~\ref{fig:vistrails} is a fully interactive 3D visualization environment
of the wildland fire simulation results, running locally on a user's machine.
These visualizations support exploration of the wildland fire through an
interactive camera, toggles for the various ground and atmospheric variables,
controls for isosurface and timing components, as well as interactive vector streamlines.

\begin{figure}[t]
\begin{center}
\includegraphics[width=7in]{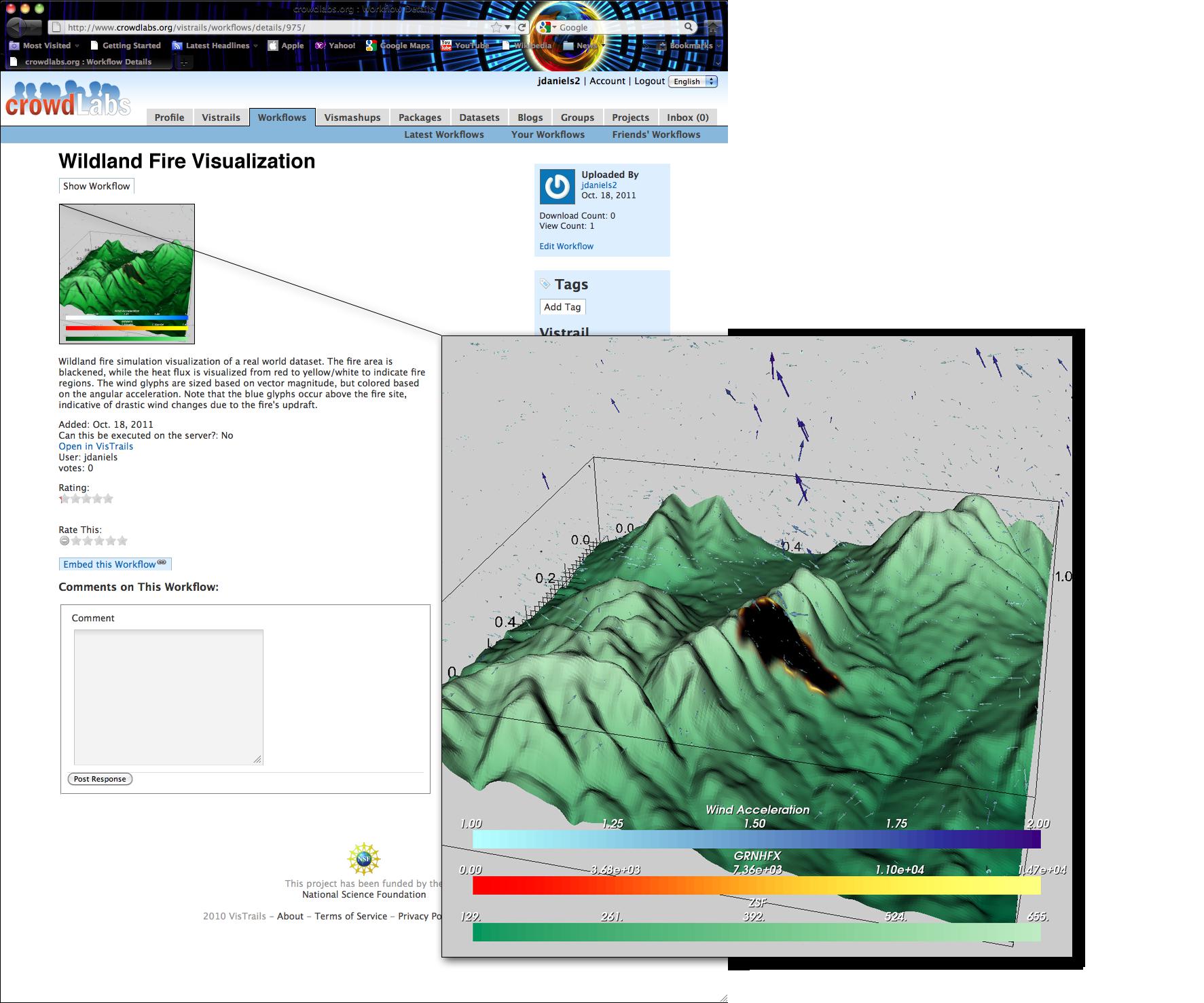}
\end{center}
\caption{Visualization in CrowdLabs web portal.}%
\label{fig:crowdlabs}%
\end{figure}

Providing web services, we improve the accessibility of the coupled WRF and
fire simulation results for educational purposes and to improve communication
and data sharing within the research community. We have setup the wildland
fire simulation user group within CrowdLabs (\citet{Mates-2011-CSA},
\url{ http://www.crowdlabs.org}). CrowdLabs is a social web site that provides
a scalable environment for collaborative analysis involving simulation
scientists, visualization experts, and other users alike. Through a web
portal, users share and discuss VisTrails workflows and VisMashup interfaces
to improve their efficacy. Fig.~\ref{fig:crowdlabs} illustrates the online
interface for CrowdLabs, as well as a shared fire visualization on the site.
It remains future work to provide 3D interactions within the web-based
visualizations, as well as develop the controls to manage simulation
executions from the site.

\section{Acknowledgements}

John Michalakes developed the support for the refined surface fire grid in
WRF. The developers would like to thank Janice Coen for providing a copy of
the Clark-Hall code with the tracer-based fire model, and Ned Patton for
providing a copy of his prototype code linking a Fortran 90 port of the fire
code with WRF. Eric Jorgensen has worked on setting up the OpenWFM.org site
and the wiki. Barry Lynn and Guy Kelman have provided data in Israel and
valuable operational considerations. Further contributions are acknowledged in
the reference paper \citet{Mandel-2011-CAF}.

\bibliographystyle{ametsoc}
\bibliography{../../references/geo,../../references/other}

\appendix

\end{document}